# STUDY OF MENTAL HEALTH AND LEARNING ENGAGEMENT DURING COVID-19 PANDEMIC BASED ON AN ELECTROENCEPHALOGRAM HEADSET

Iuliana Marin

*University Politehnica of Bucharest, Faculty of Engineering in Foreign Languages (ROMANIA)*

## Abstract

The COVID-19 pandemic and the measures which were taken had effect over the mental health of persons. In this case, students, as well as their learning process is affected. Isolation, panic, deficit of attention, confusion, depression, uncertainty and stress are words which characterize the lives of students. Virtual family care, schools and universities create new realities that can cause the evolution of severe mental conditions and study habits that need to be avoided by persons. The current paper proposes a concept that supports the performance of students by analyzing three ways of distance learning, namely text; text and illustrations, including charts; video. An electroencephalogram headset allows the detection of brainwaves and the developed web application enhances the process of distance learning. The electrodes of the headset are placed at contact with the user's head and monitor the activity of the left and right frontal regions, along with the temporal lobe. Mood, focus, stress, relaxation, engagement, excitement and interest are triggered as numerical values by using the headset. The users provide information about their daily activities, including learning and evaluation processes.

According to the study, users had the highest long-term attention while using text and illustrations, followed by watching videos. This is caused by the fact that the text contained the code for the programs which were presented in the video. Also, the users feel comfortable while using the application and they started to pay more attention to the connection between stress, health, education and well-being. The results triggered by the headset had higher values while students studied for the first-time using videos. When they wanted to remember the information, the text and illustrations way of learning was the best option. Based on the study outcomes, the instructional design can be enhanced. Moreover, the results improved as the students became more equilibrated and confident in themselves. Teachers, professors and parents are able to collaborate and enhance training, as well as support according to the data which was gathered. While studying online under lockdown, students have found the proposed solution to be good because their inner state influences their productivity while solving problems.

## 1 INTRODUCTION

The coronavirus disease pandemic had impact on the psychosocial life aspects of every person and determined the occurrence of mental health problems for a part of the individuals due to lockdown. Students are also affected due to this situation. In China, at the beginning of the pandemic, it was observed that online public education and psychological counselling have helped individuals to eliminate some of the negative mental health aspects caused by the COVID-19 pandemic [1]. In Australia and New Zealand, the request of mental health services increased by 25% due to the COVID-19 context and support was offered to them via phone, video, as well as by providing more beds in hospitals in order to treat them [2].

Mental health problems impact directly and indirectly the persons who are ill, like the ones with cardiovascular diseases. Their situations influence the lifestyle and health of their families. Rural and remote settings are in need of support services, including the post-COVID era [3]. During and after this pandemic, food security availability, access, utilization and stability have to be overviewed, because disruptions cause negative effects [4-7].

The learning process is greatly affected by the pandemic and isolation, panic, deficit of attention, confusion, depression, uncertainty and stress are words which characterize the states through which the students have to go during such times. 7143 Chinese students have reported the same daily life characteristics after answering a questionnaire [8]. A Psychology teacher from Hubei, China, affirmed that online physical education and psychology classes were done online in order to keep the students and professors physically and mentally well [9]. This was caused by the need to avoid the occurrence of severe mental conditions and bad study habits.

The current paper presents how the best learning strategy can be chosen when studying online by analyzing the results coming from an electroencephalogram headset. The brainwaves triggered by the headset electrodes and are interpreted numerically in terms of focus, stress, relaxation, engagement, excitement and interest, along with the mood of the user. The values are entered by the user via a developed web application. In the same application, information about daily activities, learning and evaluation are also provided by the user. Three ways of studying are tested as the user has the headset placed on his/her head, namely text; text and illustrations, including charts; and video. In Section 2 is outlined the proposed system which was designed to make the students more aware of their learning process engagement and inner state during the time of the COVID-19 pandemic, based on the results obtained by using an electroencephalogram (EEG) headset. Section 3 describes the results which were triggered after testing the proposed system by using three kinds of learning materials, namely text; text and illustrations; and video. The last section presents the conclusions.

## 2   METHODOLOGY

Brain-computer interface (BCI) systems use neural activity to control signals, but this is only possible after training specific cortical activity patterns [10]. Computer assisted learning is characterized by two concepts, respectively flow zone and shaping [10]. The psychologist Csikszentmihalyi introduced the notion of flow zone, meaning that the difficulty of a task is in balance with the individual's capability to be engaged without getting stressed or bored [11]. The concept of shaping was identified by a researcher, Skinner, and it represents the repetitive process through which performance is refined [12]. The individual will gradually perform complex tasks with high precision.

Neurophysiological signals triggered by an EEG based learning environment were tested in Germany and it resulted that students can adapt better to the learning content, improving in this way their efforts and the context did lead to good results [13]. An EEG headset was found to be useful to determine if there exists the need of medical intervention for the students who suffer due to anxiety [14]. Along with the EGG data, behavior and respiratory data played a key role. Breathing deeply enhances attention and proper brain electrical activity modulation which is required during the learning process.

The paper's system is dedicated to analyze the student performance during the online learning process, while being exposed to three ways of notion presentation, namely text; text and illustrations, including charts; and video. The collaboration between teachers, students and parents is desired in order to improve their education and well-being.

The gadgets which are used in the system are an Emotiv Insight headset [15], along with a personal computer and a mobile phone which are owned by the student, plus the system server, as illustrated in Fig. 1. The headset exchanges data with the student's personal computer and mobile phone via Bluetooth. The EmotivBCI [16] program is installed on the student's personal computer and it helps the user to place the headset for functioning in optimal conditions by analyzing the contact quality between the EEG electrodes and the user's head. The electrodes can get dry and cause a lower signal to noise ratio that would trigger a low prediction accuracy. Contact lens fluid or saline can be put on the electrodes for improved results.

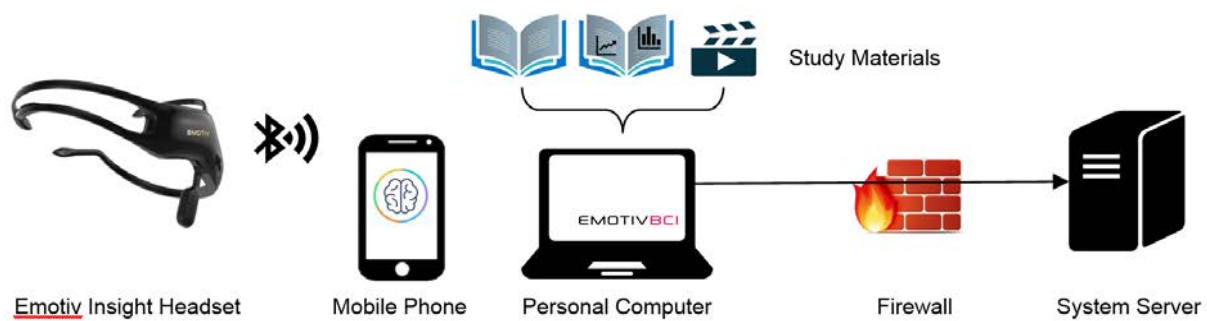

*Figure 1. System architecture*

The MyEmotiv mobile application determines based on the brainwaves which are the numerical values for focus, stress, relaxation, engagement, excitement and interest. The values are introduced manually into the system's web application, along with data about daily activities, and their learning and evaluation processes.

The student has access to different study materials that can involve just text; text and illustrations, but also videos, while using the Emotiv Insight headset. The data transfer between the user's personal computer/mobile phone and the system server is done via a route that includes a firewall to define rules which determine forbidden inbound and outbound traffic that passes through the ports of the system.

## 3    RESULTS

The Emotiv Insight headset has five EEG electrodes which trigger the brainwaves that can be visualized using the EmotivBCI program. The five channels, respectively AF3, T7, Pz, T8 and AF4, can be seen in Fig. 2.

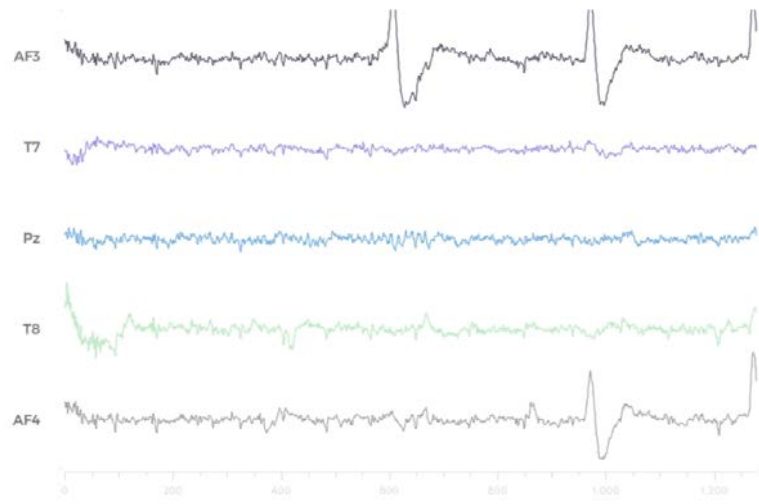

Figure 2. Channels of the EEG waves

The headset has two electrodes which are put on the user's forehead (Fig. 3). The shorter arm has the AF3 electrode which monitors the left frontal region activity, while to longer arm's AF4 electrode observes the right frontal region activity. The T7 and T8 electrodes are placed near the ear and at the back of the user's head. Both of them monitor the temporal lobe activity. The Pz electrode resides on the top of the head and it ensures the proper fitting of the headset.

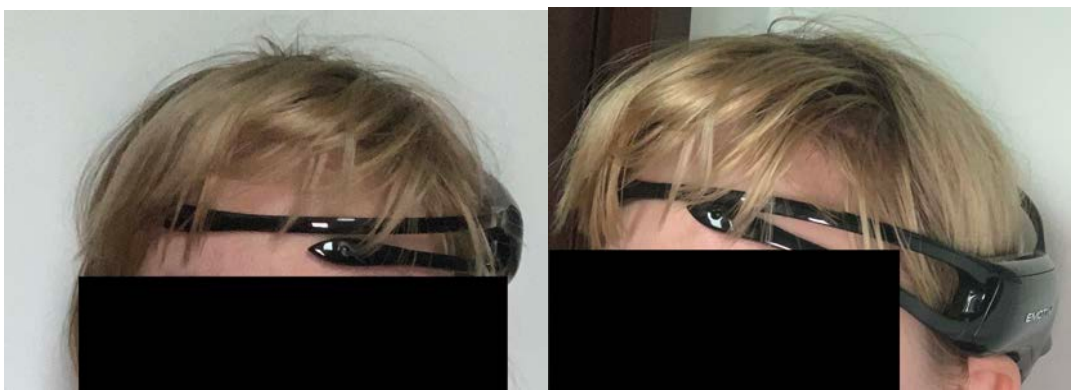

Figure 3. Headset setup for testing the system

The Centre for Neuro Skills stated that the left frontal lobe manages the control of movements. The right frontal lobe is responsible for the non-verbal abilities [17]. The temporal lobe ensures auditory perception, visual input, verbal material categorization and comprehension, memory for long term, personality [18].

The student has the MyEmotiv application [19] installed on his/her Android or iPhone mobile phone and data regarding focus, stress, relaxation, engagement, excitement and interest are given in numerical format while he/she has the headset on and studies the provided learning materials. The materials can have just text; text and illustrations; and videos. The EmotivBCI program can also determine the user's mood. The program is installed on the student's personal computer. All the student profiles are saved in the cloud of Emotiv. The student can type the triggered numerical values in the system's web interface, along with data about daily activities, learning and evaluation processes. The input values are saved at the system server side for further analysis.

The results triggered after testing the system on 10 students, showed that information is well received firstly when the learning materials are presented in video format (100%). In contrast, the long-term attention is kept when the learning materials contain text and illustrations (80%), followed by watching videos (70%). The students opted for text and illustrations due to the code of the software programs which are just illustrated in the videos. The triggered values can be used in order to enhance the design of the instructional materials which are provided to the students. Teachers, professors and parents can use the triggered values and influence positively the training process by collaborating between them.

The students were very interested in the results which are triggered by the MyEmotiv application, in what regards their focus, stress, relaxation, engagement, excitement and interest. In this way, they started to pay more attention to their state and this action had as outcome a decreased level of stress over a period of time, an improvement of their mental state, as well as in their overall distance learning educational process during the COVID-19 pandemic lockdown. The students confessed that after being conscious of the changes that occur in their body as they study, they tried to be more equilibrated and confident in themselves. As an outcome of this decision, the inner state of the students did lead to an increased productivity while solving problems.

## 4   CONCLUSIONS

The proposed solution would help the students while learning, as well as it would avoid overburdening the health care system due to the unpleasant outcomes related to mental health which are caused by the COVID-19 pandemic. The solution would also be useful for students with learning disabilities. Students who suffer due to test anxiety can be supported, because the system will trigger their state and their professors, and their parents can interfere to manage the situation.

The current paper draws attention upon the changes which occur inside the world of students, during the COVID-19 pandemic. The obtained results that were obtained using the Emotiv Insight EEG headset have showed that a special attention should be given to the students who develop themselves during lockdown. Their teachers, professors and parents can use the triggered results of the system server in the management of their teaching style. The learning materials should be more practical and be delivered in video format, as well as in text format by having solved examples. At the end of the test, the students started to be aware how their inner state influences greatly their learning process. Their daily activities, including the learning and evaluation processes are key elements in growing up healthy, satisfied and productive individuals.

## ACKNOWLEDGEMENT

The article was financed by the University Politehnica of Bucharest, Romania, through the project "Inginer în Europa", in online system, registered at MEC under no. 457/GP/06.08.2020.